\documentclass[amsfonts,showpacs,showkeys,floatfix, superscriptaddress]{revtex4}
\usepackage{graphicx}
\usepackage{dcolumn}
\begin{document}

\title{Inverse problem of variational calculus  for nonlinear evolution equations}
\author{Sk. Golam Ali, B. Talukdar}
\email{binoy123@sancharnet.in} 

\affiliation{Department of Physics, Visva-Bharati University, Santiniketan 731235, India}
\author{U. Das}
\affiliation{Department of Physics, Abhedananda Mahavidyalaya, Sainthia 731234, India}

\begin{abstract}
We couple a nonlinear evolution equation with an associated one and derive the action principle. 
This allows us to write the Lagrangian
density of the system in terms of the original field variables rather than 
Casimir potentials.
We find that the corresponding Hamiltonian density provides a natural basis to recast the pair of equations in the canonical
form. Amongst the case studies presented the KdV and modified KdV pairs exhibit bi-Hamiltonian structure and allow one to realize
the associated fields in physical terms
\end{abstract}

\maketitle
\noindent{\bf{1. \,\,Introduction}}
\vskip 0.3cm

In the calculus of variations one is concerned with two types of problems,  namely, the direct and inverse problems. 
The direct problem is essentially the conventional one in which one first assigns a Lagrangian and then computes the 
 equation of motion through Euler-Lagrange equations. As opposed to this, the  inverse problem begins with the equation 
 of motion and then constructs a Lagrangian consistent with the variational principle \cite{Santilli}. The object of the 
 present work is to derive an uncomplicated method for the Lagrangian representation  of nonlinear evolution equations. 
 We shall see that our results for the Lagrangian densities  provide a natural basis to recast
  these equations in the Hamiltonian form \cite{Zakharov,Gardner C S}.
\par 
Studies in the Hamiltonian structure of nonlinear evolution 
equations are based on a mathematical formulation that does not make explicit
reference to Lagrangians \cite{Gardner1}. We feel that the Lagrangian formulation of these
equations should be quite interesting because Lagrangian densities, via the 
Legendre map, will give us a direct route to construct the expressions for 
Hamiltonian densities that characterize the equation of Zakharov, Faddeev 
 \cite{Zakharov} and Gardner \cite{Gardner C S}. To gain  some weightage for the 
 physical and mathematical motivation of our work we  proceed by noting the following. 
\par Let $P[v]=P(x, v^{(n)})\,\,\in\,\,{\cal A}^r$ be an $r$-tuple of differentiable function. 
The Fr\'{e}chet derivative of $P$ is the differential operator $D_P$ : ${\cal A}^q\rightarrow {\cal A}^r$ defined by

$$D_P(Q)={d\over{d\in}}\arrowvert_{\in=0}\,P[v+\epsilon Q[v]]\eqno(1)$$ 
for any $Q\in{\cal A}^q$. If ${\cal D}=\sum\limits_{J}P[u]D_J$, $P_J\in{\cal A}$ is a differential operator, its adjoint $D^\star$ is 
given by 
$$D^\star=\sum_{J}(-D)_J. \,\,P_J.\eqno(2)$$
Helmholtz theorem for inverse variational problem \cite{Olver P J} asserts that any nonlinear evolution equation $u_t=P[u]$
will have a Lagrangian representation only if $D_P$ is self-adjoint. When the self-adjointness is guaranteed, a Lagrangian 
density ${\cal L}$ for $P[u]$ can be explicitly constructed using the homotopy formula 
${\cal L}[v]= \int^1_0 vP[\lambda v]d\lambda$. A single evolution equation is never an Euler-Lagrange expression.
One common trick to put a single evolution equation into the variational form is to replace u by a potential function $w$ with
$u=-w_x$. This yields $w_{xt}=P[w_x]$.The function $w$ is often called the Casimir potential. For many nonlinear evolution
equations the Fr\'{e}chet derivative of $P[w_x]$  is self adjoint, while that of $P[u]
$ is
not. In view of this, the  Lagrangian densities for single evolution equations are 
always written in terms of partial derivatives of $w$. There are some  equations for which even $P[w_x]$ is not self-adjoint. 
These are often referred to as non-Lagrangian. The well-known Burgers equations and nonlinear evolution 
equations with nonlinear dispersive terms \cite{Rosenau P} serve as typical examples  of  
 non-Lagrangian equations.  
 \par
 Keeping the above in view we shall work out a Lagrangian representation without taking recourse to the 
 use of Casimir potentials. In section 2 we  introduce a suitable associated equation for an additional
field variable $v(x,t)$ and use it to write the action principle for any nonlinear evolution equation. We express the 
Lagrangian density  in terms of $u(x,t)$, $v(x,t)$  and their partial derivatives. We show in section 3 that the coupled 
set of equations as introduced by us form a Hamiltonian system. We also cite examples in which the associated fields admit
simple physical realization. In section 4 we summarize our outlook on the present work and make some concluding remarks.
\vskip 0.5cm
\noindent{\bf{2. \,\,Lagrangian representation}}
\vskip 0.3cm

The basic philosophy we use here to construct expressions of Lagrangian  densities of nonlinear evolution equations has  a rather old root 
in the classical-mechanics literature. For example, as early as 1931, Bateman  \cite{Bateman H} allowed for an 
additional degree of freedom to bring the equation of motion for the damped harmonic
 oscillator within the framework of action principle. 
The Bateman Lagrangian 
$$ L={\dot x}{\dot y}+{\gamma\over2}(x\dot y-y\dot x)-\omega^2xy \eqno(3)$$
for the one-dimensional linearly damped harmonic oscillator
$$\ddot x+\gamma \dot x+\omega^2x=0,\,\,\,x=x(t)\eqno(4)$$
has a 'mirror image' equation
$$\ddot y-\gamma \dot y+\omega^2y=0,\,\,\,y=y(t)\eqno(5)$$ 
for the associated coordinate $y(t)$. Here $\gamma$ is the co-efficient of 
friction and $\omega$, the natural frequency of the oscillator. The overdots
stand for derivatives with respect to time t. Understandably, the 
complementary equation in (5) represents a physical system which absorbs energy
dissipated  in the first. Interestingly, Bateman \cite{Bateman H} regarded a dissipative system as physically incomplete such
that one needs to bring in an additional equation to derive the original one from an action principle. Thus it remains a real 
curiosity to envisage a similar study in the context of classical fields and look for the Lagrangian representation of nonlinear
evolution equations.
\par
Any nonlinear evolution equation that has at least one conserved density 
${\rho[u]}$ can be written in the form 
$$ u_t+{\partial\over \partial x}{\rho[u]}=0,\,\,\,u=u(x,\,t).\eqno(6)$$
This single evolution equation is non-Lagrangian. When written in terms of the 
Casimir potential, the equation resulting from (6) may be either  Lagrangian or 
non-Lagrangian.
However, we can make use of an elementary lemma to get a Lagrangian representation of $(6)$.
\vskip 0.5cm
\noindent{\bf Lemma 1.} There exists a prolongation of $(6)$ into another equation
$$v_t+{\delta\over\delta u}(\rho [u]v_x)=0,\,\,\,v=v(x,\,t)\eqno(7)$$
with the variational derivative
$${\delta\over\delta u}=\sum_{k=0}^{n}(-1)^k{\partial^k\over\partial x^k}
{\partial\over \partial u_{kx}},\,\,\,\,\,u_{kx}={\partial^k
 u\over\partial x^k} \eqno(8)$$
such that the system of equations follows from the action principle 
$$\delta \int {\cal L}\,\,dx dt=0.\eqno(9)$$
Here ${\cal L}$ stands for the Lagrangian density.
\vskip 0.5cm
\noindent{\bf Proof.}  For a direct proof of the lemma let us introduce ${\cal L}$ in the form
$$ {\cal L}={1\over 2}(vu_t-uv_t)-\rho[u]v_x.\eqno(10)$$ 
 From (9) and (10) we obtain the Euler-Lagrange equations 
$${d\over dt}(\frac{\partial {\cal L}}{\partial v_t})-
\frac{\delta {\cal L}}{\delta v}=0\eqno(11)$$
and $${d\over dt}(\frac{\partial {\cal L}}{\partial u_t})- \frac{\delta {\cal L}}{\delta u}=0.\eqno(12)$$ 
Using (10) in (11) and (12) we obtain (6) and (7) respectively.
\vskip 0.5cm 
\par
We shall now consider two examples of physical interest and apply the rule  in (7) to construct the associated equations. 
We first focus our attention on
the Korteweg-de Vries (KdV) equation
$$u_t+6uu_x+u_{3x}=0.\eqno(13)$$
The KdV equation represents the prototypical nonlinear evolution equation that
was first solved by the inverse spectral transform method \cite{Gardner2}. From (6) 
and (13) we see that for this equation 
$$\rho[u]=3u^2+u_{2x}.\eqno(14)$$
Using (14) in (7) we get the associated equation
$$v_t+6uv_x+v_{3x}=0.\eqno(15)$$
The corresponding Lagrangian density as obtained from (10) reads 
$${\cal L}={1\over 2}(vu_t-uv_t)-(3u^2+u_{2x})v_x.\eqno(16)$$
It is easy to verify that (16), when substituted in (11), reproduces the KdV equation.
The second example of our interest is the so-called modified KdV (mKdV) equation given by
$$ u_t+6u^2u_x+u_{3x}=0.\eqno(17)$$
Equation $(13)$ and $(17)$ are connected by Miura transform and as with $(13)$, $(17)$ can also be solved by the inverse spectral
method\cite{F. Calogero}. The mKdV equation appears in a number of applicative contexts including 
description of Alfv\'{e}n waves in a collisionless plasma. The associated
equation for $(17)$ is obtained in the form 
$$ v_t+6u^2v_x+v_{3x}=0\eqno(18)$$ 
with the Lagrangian density given again by $(10)$. For the mKdV equation 
$$\rho[u]=2u^3+u_{2x}.\eqno(19)$$
Results similar to those  for KdV and mKdV equations can also be written for other nonlinear evolution  equations which can 
be expressed in the form (6). 
We give in Table I, the results for  a number of evolution equations which are often believed to be non-Lagrangian.
\begin{widetext}
{\bf Table I.} Associated equations for a few important nonlinear  evolution equations
\begin{center}
\begin{tabular}{|c|c|c|}
\hline
Evolution equation&Conserved density&Associated equation\\
\hline
Burgers2: &&\\
$u_t-u_{2x}-2uu_{x}=0$&$-(u_x+u^2)$&$v_t+v_{2x}-2uv_{x}=0$        \\ 

Burgers3:&&\\
$u_t-u_{3x}-3u^2u_{x}-$&$-(u_{2x}+u^3+3uu_x)$&$v_t+3uv_{2x}-3u^2v_x$ \\  
$3uu_{2x}-3u_x^2=0$&&$-v_{3x}=0$\\
KdV-Burgers:&&\\
$u_t+uu_x-\nu u_{2x}+$&${1\over 2}u^2-\nu u_x+\mu u_{2x}$&$v_t+uv_x+\nu v_{2x}+$  \\
$\mu u_{3x}=0$&&$\mu v_{3x}=0$\\
FNE3:&&\\
$u_t+3u^2u_x+6u_xu_{2x}+$&$u^3+2u^2_x+2u u_{2x}$&$v_t+3u^2v_x+ 2uv_{3x} =0$   \\
$2uu_{3x}=0$&&\\
FNE5:&&\\
$u_t+\beta_1(u^2)_x+\beta_2(u^2)_{3x}$&$\beta_1 u^2+\beta_2(u^2)_{2x}+\beta_3(u^2)_{4x}$&$v_t+2\beta_1uv_x+2\beta_2uv_{3x}$     \\
$+\beta_3(u^2)_{5x}=0$& &$+2\beta_3uv_{5x}=0$\\ 
\hline
\end{tabular}
\end{center}
\end{widetext}
The first two equations in the Table are due to Burgers. These are dissipative and do not support soliton solutions. 
However, both of them are useful in the study of acoustics and shock waves \cite{Crighton}. 
In the recent past two of us \cite{Talukdar} studied the equations in the Burgers hierarchy and sought a 
Lagrangian representation in which the appropriate equations were expressed in terms of Casimir potentials.
 However, the results represented here have the obvious virtue of simplicity and directness.The compound KdV-Burgers equation \cite{Parkes}  
describes wave propagation in which the effects of nonlinearity, dissipation and  dispersion are all present.
 We believe that the Lagrangian representation of these equations using  associated equations is quite interesting. The KdV
 and mKdV equations are quasi-linear in the sense that the dispersive behaviour of the solution of each equation is 
 governed by a linear term giving the order of the equation. In contrast to this, the third and fifth  order equations
 (FNE3) and (FNE5) \cite{Rosenau P} in the table are nonlinear partial differential equations with nonlinear dispersive terms. 
These are, therefore, fully nonlinear evolution (FNE) equations. The solitary wave solutions of these equations have a compact
support. To the best of our knowledge no Lagrangian representations of FNE3
and  FNE5 have yet been found. There have,however, been attempts \cite{Cooper F} to introduce Lagrangian system of 
FNE equations which support compacton solutions.
\vskip 0.5cm
\noindent{\bf{3. \,\,Canonical structure}}
\vskip 0.3cm
Zakharov and Faddeev \cite{Zakharov} developed the Hamiltonian approach to integrability of nonlinear evolution equations in 
one spatial and one temporal $(1+1)$ dimensions and Gardner \cite{Gardner C S}, in particular, interpreted the KdV equation as a 
completely integrable Hamiltonian system with $\partial_x$ as the relevant Hamiltonian operator. In this context we introduce
the following lemma for the Hamiltonian structure of the coupled equations introduced by us.
\vskip 0.5cm
\noindent{\bf Lemma 2.} The Hamiltonian density ${\cal H}$ constructed from $(10)$, can be used to express $(6)$ and $(7)$  in the 
Hamiltonian form
$$ {\eta}_t={ J}\frac{\delta {\cal H}}{\delta { \eta}}\eqno(20)$$ with 
$${ \eta}=\left(\begin{array}{c}
u\\
v
\end{array}\right)\,\,\eqno(21a)$$
 and the  symplectic matrix
$${ J}=\left(\begin{array}{cc}
0&1\\
-1&0
\end{array}\right)\,\,.\eqno(21b)$$
\vskip 0.5cm
\noindent{\bf Proof.}  Using the Legendre map the Hamiltonian density for the Lagrangian in $(10)$ is obtained as 
$${\cal H}=\rho[u]v_x.\eqno(22)$$
From $(20)$,$(21)$ and $(22)$, equations in $(6)$ and $(7)$ follow immediately
\vskip 0.5cm
\par
A significant development in the Hamiltonian theory is due to Magri \cite{F. Magri} who realized that completely integrable 
Hamiltonian system have an additional structure. They are bi-Hamiltonian systems, i.e.,they are Hamiltonian with respect to two different compatible Hamiltonian operators. We have found that
both KdV and mKdV can be recast in the bi-Hamiltonian form 
$$ {\eta}_t={ J_1}\frac{\delta {{\cal H}_2}}{\delta { \eta}}={ J_2}\frac{\delta {{\cal H}_1}}{\delta { \eta}}.\eqno(23)$$
The appropriate results for Hamiltonian operators and Hamiltonian densities are given by
$$J_1^{KdV/mKdV}=J,\eqno(24)$$
$$J_2^{KdV}=\left(\begin{array}{cc}
0&(\partial^3_x+2u\partial_x+2\partial_xu)\partial_x^{-1}\\
-(\partial^2_x+4u)&2v_x\partial^{-1}_x
\end{array}\right),\eqno(25)$$
$$J_2^{mKdV}=\left(\begin{array}{cc}
 0&{(\partial^3_x+2u^2\partial_x+{4\over3}\partial_xu^2)\partial_x^{-1}}\\
-(\partial^2_x+4u^2)&2uv_x\partial^{-1}_x
 \end{array}\right),\eqno(26)$$
 $${\cal H}^{KdV/mKdV}_1=uv_x,\eqno(27)$$
 $${\cal H}^{KdV}_2=(3u^2+u_{2x})v_x\eqno(28)$$ and
 $${\cal H}_2^{mKdV}=(2u^3+u_{2x})v_x.\eqno(29)$$
 \par
In solving the inverse variational problem for nonlinear evolution 
equations we coupled the field variable $u(x,t)$ of some given equation with
the field variable $v(x,t)$ of an associated equation such that the system 
follows from the action principle with a prescribed form of the Lagrangian 
density as given in $(10)$. Admittedly, one of our tasks in this work will be to 
study the nature of $v(x,t)$ when the original field variable $u(x,t)$ admits 
simple physical realization. Keeping this in view,  we focus our attention on the  KdV and mKdV equations which support 
soliton solutions. It is well-known that the solutions of $(13)$ and $(17)$ are given in the general form
$$u^{KdV}(x,t)=A(k)sech^2(kx-4k^3t)\eqno(30)$$
 $$u^{mKdV}(x,t)=B(k)sech(kx-k^3t),\eqno(31)$$
where $k$ is the wave number for a single bound-state energy of the potential that characterizes the spectral 
problem in solving the evolution equation \cite{Gardner2}. Here $A(k)$ and $B(k)$ represent the amplitudes of the  bright solitons
in $(30)$ and $(31)$. Understandably, both solutions $u^{KdV}(x,t)$ and $u^{mKdV}(x,t)$ are centered at $x=0$. The KdV solution 
moves to the right with speed $4k^2$ while the mKdV solution moves in the same direction with speed $k^2$ only. One can 
verify that the associated fields corresponding $u^{KdV}(x,t)$ and $u^{mKdV}(x,t)$ are given by the dark soliton solutions
$$v^{KdV/mKdV}(x,t)=\tanh(kx-4k^3t).\eqno(32)$$
From $(30)$ and $(32)$ it is clear that both soliton solutions of the KdV pair move with equal speed. As opposed to this, 
comparison of $(31)$ and $(32)$ reveals that the speed of the dark soliton solution for the mKdV pair is four times 
the speed of the bright soliton solution. This is not immediately clear to us and deserves extensive numerical study for further
clarification.
\vskip 0.5cm
\noindent{\bf{4. \,\,Concluding remarks}}
\vskip 0.3cm
Nonlinear evolution equations are not directly amenable to Lagrangian representation. 
We have established that if (6) does not admit a direct analytic or Lagrangian representation, then there exists an auxiliary or 
associated field which helps us treat the original evolution 
equation within the framework of the action principle. The method is quite general and works for both integrable and nonintegrable 
equations. We have dealt with equations in which the effects of nonlinearity, dissipation and dispersion are all present.
\par
It is a well-known belief that there is no exact method for applying variational principle to dissipative systems. In view of this, studies
in Lagrangian and Hamiltonian mechanics of nonconservative systems are still regarded as an interesting curiosity. In the context of 
point mechanics Riewe \cite{Riewe} used the method of fractional calculus to write a Lagrangian for $(4)$ without taking help of the
additional equation in $(5)$. Kaup and Malomed \cite{Kaup} sought an application of the variational principle to nonlinear field equations 
involving dissipative terms. The ansatz for the Lagrangian density used by these authors is simply related to our expression for ${\cal L}$
in Lemma 1. For example, we can add the gauge term $\frac{d}{dt}\left({1\over 2}uv\right)$ + 
$\frac{d}{dx}\left(3u^2v\right)$ + $\frac{d}{dx}\left(u_{2x}v\right)$  on the right side of $(16)$ and write
$${\cal L}=v\left(u_t+6uu_x+u_{3x}\right).\eqno(33)$$

In this context we note that $(33)$ was originally suggested by Atherton and Homsy \cite{Atherton} and subsequently included as an exercise ( 5.37, page 184)  in ref. 5.
However, it appears that there is no physically founded assumptions in writing ${\cal L}$ in this form except that a Lagrangian may involve 
its own equation of motion provided one introduces a new concept of variational symmetry called the s-equivalence \cite{Hojman}. 
On the other hand, the treatment presented here is based on a fomalism that is specially intended to bring out the reasons why
nonloinear evolution equations, as such, do not follow from the action principle.
\par
The KdV-like equations support bright solitons while the associated fields  
have soliton solutions which are dark. The coupled set of equations for the bright
and dark solitons are Lagrangian although, individually, each of them is non-Lagrangian.
This observation appears to bring in some similarity with the celebrated work of Bateman \cite{Bateman H}  on the dissipative system
in particle dynamics.
\par
The appearance of dark solitons in the solutions of our coupled equations is not a strange phenomenon. In the past, 
dynamics of dark solitons induced by stimulated Raman effect in the optical fiber were explained by the use of 
KdV-Burgers equation \cite{Kivshar}. We feel that the inverse variational problem as treated here will be useful to
study Noether symmetries and even to construct exact solutions of the nonlinear evolution equations \cite{Kara}.
\vskip 0.5 cm
\noindent{\bf{Acknowledgements}}\\
\vskip 0.1 cm
{\small{This  work forms the  part  of a Research Project F.10-10/2003(SR) supported  by the University Grants Commission, 
Govt. of India. One of the authors (SGA) is thankful to the UGC, Govt. of India for a Research Fellowship. }}


\begin{thebibliography}{department}
\bibitem{Santilli} R. M. Santilli, Foundations of Theoretical Mechanics, Vol 1, The Inverse Problem in Newtonian Mechanics
(Springer - Verlag, New York, 1978 ).
\bibitem{Zakharov} V. E. Zakharov, L. D. Faddeev, Funct. Anal. Appl. {\bf 5}  18 (1971). 
\bibitem{Gardner C S} C. S. Gardner, J. Math. Phys. {\bf 12}  1948 (1971).  
\bibitem{Gardner1} C. S. Gardner, J. M. Greene, M. D. Kruskal, R. M. Miura, Commun. Pure
Appl. Math. {\bf 27} 97  (1974).  
\bibitem{Olver P J} P. J. Olver, Application of Lie group to differential equations, (Springer - Verlag, New York, 1993 ).
\bibitem{Rosenau P}  P. Rosenau, J. M. Hyman, Phys. Rev. Lett. {\bf 70}  564 (1993).
\bibitem{Bateman H}  H. Bateman, Phys. Rev. {\bf 38}  815 (1931). 
\bibitem{Gardner2} C. S. Gardner, J. M. Greene , M. D. Kruskal, R. M. Miura Phys. Rev. Lett. {\bf 19}  1095 (1967). 
\bibitem{F. Calogero}  F. Calogero,  A. Degasperis, Spectral transform and solitons, Vol 1 (North-Holland Pub. Co., Amsterdam , 1982 ). 
\bibitem{Crighton} D. G. Crighton,  Basic Nonlinear Acoustics (Frontiers in Physical Acoustics) Ed.  Sette D (North - Holland 
Pub. Co., Amsterdam, 1986 ) 
\bibitem{Talukdar} B. Talukdar, S. Ghosh,  U. Das, J. Math. Phys. {\bf 46}  043506 (2005). 
\bibitem{Parkes} E. J. Parkes, Phys. Lett. A {\bf 317}  424 (2003). 
\bibitem{Cooper F}  F. Cooper,  H. Shepard, P. Sodano, Phys.Rev. E{\bf 48} 4027  ( 1993);
  F. Cooper, J. M. Hyman, A. Khare, Phys.Rev. E {\bf 64} 026608  (2001); S. Ghosh, U. Das, B. Talukdar,
  Int. J. Theo. Phys. {\bf 44} 363  (2005 ). 
\bibitem{F. Magri}  F. Magri, J. Math. Phys. {\bf 19} 1156  (1978 ).
\bibitem{Riewe}  F. Riewe, Phys. Rev. E {\bf 55} 3581  (1997).
\bibitem{Kaup} D. J. Kaup, B. A. Malomed, Physica D {\bf 87} 155  (1995).
\bibitem{Atherton} R. W. Atherton and G. M. Homsy, Stud. Appl. Math.  {\bf 54} 31  (1975).
\bibitem{Hojman}  S. Hojman, J. Phys. A {\bf 17} 2399  (1984).
\bibitem{Kivshar} Y. S. Kivshar, Phys. Rev. A {\bf 42} 1757  (1990). 
\bibitem{Kara} A. H. Kara, C. M. Kalique, J. Phys. A {\bf 38} 4629 (2005) . 
\end{thebibliography}
\end{document}